# Mechanics of Mineralized Collagen Fibrils upon Transient Loads

Mario Milazzo[1,2], Gang Seob Jung[1], Serena Danti[1,2,3], Markus J. Buehler[1,4*]

[1] Laboratory for Atomistic and Molecular Mechanics (LAMM), Massachusetts Institute of Technology, 77 Massachusetts Ave, Cambridge, MA 02139, USA
[2] The BioRobotics Institute, Scuola Superiore Sant'Anna, Viale Rinaldo Piaggio 34, 56025 Pontedera (PI), Italy
[3] Dept. of Civil and Industrial Engineering, University of Pisa, Largo L. Lazzarino 2, 56122 Pisa, Italy
[4] Center for Computational Science and Engineering, Schwarzman College of Computing, Massachusetts Institute of Technology, 77 Massachusetts Ave, Cambridge, MA 02139, USA

*Correspondence should be addressed to:

    Prof. Markus J. Buehler
    Laboratory for Atomistic and Molecular Mechanics (LAMM)
    Massachusetts Institute of Technology, 77 Massachusetts Ave., Cambridge, Massachusetts 02139, USA
    mbuehler@mit.edu, +1.617.452.2750

**Abstract**
Collagen is a key structural protein in the human body, which undergoes mineralization during the formation of hard tissues. Earlier studies have described the mechanical behavior of bone at different scales highlighting material features across hierarchical structures. Here we present a study that aims to understand the mechanical properties of mineralized collagen fibrils upon tensile/compressive transient loads, investigating how the kinetic energy propagates and it is dissipated at the molecular scale, thus filling a gap of knowledge in this area. These specific features are the mechanisms that Nature has developed to passively dissipate stress and prevent structural failures. In addition to the mechanical properties of the mineralized fibrils, we observe distinct nanomechanical behaviors for the two regions (*i.e.*, overlap and gap) of the *D*-period to highlight the effect of the mineralization. We notice decreasing trends for both wave speeds and Young's moduli over input velocity with a marked strengthening effect in the gap region due to the accumulation of the hydroxyapatite. In contrast, the dissipative behavior is not affected by either loading conditions or the mineral percentage, showing a stronger damping effect upon faster inputs compatible to the bone behavior at the macroscale. Our results offers insights on the energy dissipative behavior of mineralized collagen composites to design and characterize bioinspired composites for replacement devices (*e.g.,* prostheses for sound transmission or conduction) or optimized structures able to bear transient loads, *e.g.,* impact, fatigue, in structural applications.

Bone is the main constituent of the skeletal system in humans and animals, and it has interesting mechanical properties due to its peculiar hierarchical structure. Bone serves as a scaffold for musculature and protects organs and soft tissues.[1] In addition to being a passive frame for other systems, bone is also an active player in the auditory apparatus, as it is able to conduct sounds as occurs in the temporal bones while the ossicular chain enables sound transmission from the eardrum to the cochlea. Moreover, the ossicular chain not only provides material continuity throughout the internal structures, but also acts as a high-pass filter for sounds in noisy environments.[2]

These distinct roles possessed by bone (*i.e.*, protecting/scaffolding soft tissues and transferring/dissipating energy) has continuously pushed researchers to unveil the intrinsic mechanisms of such an interesting material. A large number of studies have been carried out at the macroscale to highlight mechanical properties (*e.g.*, strength, toughness) upon transient and regular loads, discovering responses highly dependent on directionality and bone type.[3–6] At the same time, researchers have investigated the properties of bone hierarchical structure, depicted in **Figure 1A**, from the macroscale (length ~cm) to the atomistic scale (length ~Å), which is mainly composed of collagen (COL) and a mineral component similar to hydroxyapatite (HA).[7–9] Although HA differs from the native material for other impurities, from the Materials Science standpoint the COL/HA composite has been universally considered a basic building block for understanding bone tissue.[10–13] The arrangement of the mineral and organic components has been described in the periodical repetition of the so called *D*-period, originally defined in collagen without mineral, which is composed of two regions named overlap and gap (**Figure 1B**).[14] The gap region, 0.54*D* (≈36 nm), is the volume in which the mineral mostly fills the interspaces between collagen fibers,[15,16] and produces a heterogeneity to the material that is responsible for the piezoelectric properties of the bone.[17] Earlier studies have provided an exhaustive description of the mechanics of such a composite upon slow tensile deformation, observing a monotonic stiffening and embrittlement of the structure with the %HA that balances the ductility possessed by collagen.[16]

In view of acoustic applications, or more broadly for case studies involving transient loads, earlier studies have investigated wave propagation and energy dissipation through proteins and collagen molecules.[18–21] Recently, it has been demonstrated how collagen molecules differently respond based on load directionality and hydration state. Assessing the details, while no relevant differences have been observed upon longitudinal impulsive loads, a peculiar behavior is revealed when transversal loads are applied to the triple helix. In this case, it was estimated that wave dissipation along hydrated collagen peptides is ≈5 times higher than in dry structures.[22] Building on these earlier works, here we present an investigation on the mechanical behavior of COL/HA composite at the molecular scale upon transient loads for optimized bioinspired materials and structures to be employed in applications not exclusively limited to tissue engineering.

We investigate collagenous topologies with different weight percentages (w/w%) of HA (HA/COL 0/100, 20/80, and 40/60, hereafter recalled as 0%HA, 20%HA, and 40%HA) to describe the specific role of COL and HA in transferring and dissipating energy in bone upon tensile and compression impulsive loads (**Figure 1B**). The mechanical behavior of bone exposed to transient loads has thus far rarely been investigated at the molecular scale, especially in connection with specific macro-scale phenomena. For instance, researchers have demonstrated that vibrations and mineralization of bone have an interesting relationship, revealing how bone is able to grow and to heal when exposed to ultrasounds (*i.e.*, frequencies above 20 kHz).[23,24] However, it is still unclear how the mineralization percentage affects wave propagation and energy dissipation.

**Results and discussion**

We present an *in silico* study on the mechanical characterization of mineralized fibrils upon impulsive loads that aims at completing the description of the mechanics of the collagenous tissues at the atomistic scale. In contrast to the previous works,[16,25–29] our approach aims at studying bone and its mechanical properties by observing the propagation and dissipation of the energy from an external impulsive displacement load at the molecular level.
We use full atomistic simulations to model fibrils with different %HA (**Figure 1B**), and divide the topology in a finite number of slices, fixing the displacements of the atoms at one end, and loading the atoms at the opposite end with an impulsive displacement-based load (**Figure 2**). We perform a sensitivity analysis on the %HA and input velocity ($v_i$) in order to observe the variation of the materials' response upon slow/fast inputs. We use the axial displacements of $C_\alpha$ atoms as the main benchmark to study how energy is transmitted and dissipated along the structures. To achieve the clearest presentation of the results, we normalize the averaged axial displacements of the $C_\alpha$ atoms belonging to each slice of the model with respect to the load amplitude.
In
**Figure 3A, B**, we show the normalized averaged axial displacements of the $C_\alpha$ atoms as a function of the time, position along the model, %HA, input velocity ($v_i$), and loading conditions, normalized over the maximum value. We use both compressive and tensile loads for a complete characterization of the material: although bone structures are usually subjected to compressive stresses, tensile stress fields occasionally occur in specific loading conditions (*e.g.*, bending of the homer while handling loads).

Globally, load directionality does not severely affect the energy propagation, giving a quite similar picture in both
**Figure 3A**, **B**. However, it is possible to observe how, locally, the mechanical waves differently transit throughout the structure with local peaks noticeable with pushing loads that are, instead, dampened with tensile conditions. This different behavior may be explained by the local entanglement of the fibrils that buckle under compressive loads.

Independently of the loading conditions, the transition of the mechanical waves is severely affected by the presence of the HA that is mostly accumulated in the gap region. The presence of the mineral at the interface between the overlap and gap regions (depicted as a dashed line in
**Figure 3***C) acts as a "wall" that partially reflects the mechanical waves. This phenomenon is clearly observable for both loading conditions: we qualitatively highlight, by way of example, the traveling wave and reflected waves in* **Figure 4**
**Figure 3** (detail of
**Figure 3A** for $v_i = 50$ m·s$^{-1}$ and 20% HA) with a solid black line and red solid lines respectively. Reflected waves are generated every time the main traveling wave crosses the interface between the overlap and gap regions. An animation of **Figure 4** is provided in the Supplementary Information. We analyze the variation of kinetic energy between the overlap and gap regions at the first crossing of the interface, and we estimate an average loss percentage of 7.55%, whose values increase along with the % HA.

Moreover, we see a sharp change of slopes at 10 ps around the gap region with an increase of the ≈40% for the 20%HA and ≈45% for the 40%HA. A modest change due to the non-homogeneity of the material is also present for the non-mineralized collagen (*viz.*, ≈10%). This behavior directly influences the stiffness of the compounds and well matches the results achieved by Nair *et al.*, in which the boldest variation of the stiffness was estimated between the 0%HA and the 20%HA (+ ≈80%) with a modest difference between the 20%HA and 40%HA (+ ≈30%).[16] Moreover, the direct relationship between the different percentages of HA and the increase of the stiffness of the material, may offer insights on the evolution of the cellular response in scaffolds when the local microenvironment changes due to mineralization,[30–33] as well as on bone remodeling in larger scale applications.[8,34–37] We further carry out a sensitivity analysis on the loading-rate conditions by varying the input velocity ($v_i$) from 50 to 1000 m·s$^{-1}$. **Figure 5** summarizes the results achieved after post-processing the displacement field of $C_\alpha$ atoms.

We observe a global decrease of the wave speeds, and consequently of the Young's moduli, with the increase of $v_i$, and this is in contrast with previous studies in which researchers have observed the opposite behavior for single peptides.[20,38] In our opinion, the shift of this behavior is to be attributed to the hydrogen bond (H-bond) network that connects the peptides within the fibrils, which is more complex and extensive than that of the single triple helix. We believe that, in our case, H-bonds are more easily disturbed by high speeds and present more flexibility, thus softening the overall structure. An in-depth work encompassing a dedicated model at different observation scales is still missing, and would allow the biomechanics of this specific aspect to be fully understood.

We estimate also the acoustic impedance of the 40%HA topology that is ~4×10$^6$ kg·m$^2$·s$^{-1}$ for both compressive/tensile loading conditions. This numerical value is below the bone's (~7.5×10$^6$ kg·m$^2$·s$^{-1}$),[39] but this is due to fact that bone has a higher mineralization, up to 70%,[40] and our model is an approximation of the native tissue, being able to take into account only the intrafibrillar mineralization,[16,29] leaving out the extrafibrillar mineralization.[41]

In **Figure 6**, we provide a depiction of the estimated Young's moduli over the %HA. We compare the results from our study, identified as "Impulsive load model", with the outcomes from a previous work using atomistic simulations,[16] and two analytic models provided by Halpin-Tsai (HT) and Gao.[42,43] We show with star markers the Young's moduli across the %HA and input velocities from our investigation. The solid red line depicts the quadratic fit of such data, which has an intercept ($E_m$ = 2.72 GPa) that corresponds to the elasticity of the composite matrix (aka pure collagen).

The HT model also takes the $E_m$ into account (Eq. $E = E_m \frac{1 + A \times B \times \%HA}{1 - B \times \%HA}$  Eq. 6): the solid blue line originates by $E_m$ = 2.72 GPa from impulsive loads, while the blue dashed line employs $E_m$ = 0.5 GPa as it was estimated in [16], in which the topologies are subjected to quasi-static loads.

The model proposed by Gao predicts Young's moduli that are far below the other estimations. This difference is mainly due to the dependence of the model on the shear modulus of collagen ($G_p$ in Eq. $E = \left[\frac{4(1-\%HA)}{\%HA^2 \times \gamma^2 \times G_p} + \frac{1}{\%HA \times E_p}\right]^{-1}$  Eq. 8) that is lower than the mineral's by several orders of magnitude. The results from the atomistic simulations in [16] (blue circled markers) better agree with Gao's but are far below the HT curve at slow deformations (blue dashed line).

In contrast, our results fall between the two curves related to the HT models. The high Young's moduli and the discrepancy with respect to earlier studies is because the material is loaded faster, triggering the loading-rate dependency of collagen.[18,22,44] At the same time, our simulation results are below the curve of the HT model that is tuned for impulsive loads through the $E_m$, confirming the overestimation of the HT model that was already observed in [16].

Lastly, a further reason for the mismatch between our estimated Young's moduli with the cited analytical models may concern the employment of Eq. $v = \sqrt{\frac{E}{\rho}}$  Eq. 2, which correlates the wave speed with the elasticity of the material through its density. This formula gives a reliable representation of the longitudinal vibration phenomenon at the macroscale, but it cannot take into account any secondary three-dimensional effect related to molecular deformations. However, despite these limitations, Eq. $v = \sqrt{\frac{E}{\rho}}$  Eq. 2 has extensively been used, as a first approximation, to post-process atomistic simulation data (*e.g.*, [20,22]).

**Figure 7** depicts the trend of the relaxation time (τ), along the *x*-axis, as a function of the input velocity $v_i$. Independently of the loading conditions and the %HA, all the topologies show a monotonic decrease of τ over $v_i$ up to 500 m·s$^{-1}$ when we notice a substantial saturation. This behavior is different than that of the almost homogenous and isotropic materials (*e.g.*, steel, polycarbonate) that present a dissipative behavior quite constant and independent from the input velocity.[45] Moreover, this feature well matches with the capability of the bone at the macro scale to efficiently dampen external loads. In a previous study, we described the mechanical behavior of a single peptide, analyzing also the damping mechanisms of the triple helix upon an impulsive compressive load with input velocity equal to 100 m·s$^{-1}$. The relaxation time measured for a single molecule is ≈3 times longer than the one estimated in the present study (100 ps *vs.* 34.3 ps).[22] The dissipation behavior is mainly due to the organic component of the material, which carries most of the deformation (especially in the overlap region) in contrast to the mineral component that stores most of the stress.[16,21] Specifically, collagen peptides upon tensile loads show the tendency to uncoil,[46] and eventually slide on the HA minerals:[47] this phenomenon also observed at a larger scale[48] is one reason for the dissipation. Although our model could allow the observation of the uncoiling mechanisms of collagen, it is not possible to observe them in our simulations since, as described in a previous study,[38] the triple helices lose their configuration upon low-velocity inputs, losing hydrogen bond by rotation, below 50 m·s$^{-1}$. However, due to the marginal differences observed upon compressive/tensile loads on the relaxation time, summarized in **Figure 7A**, we believe that this is not the main factor responsible for dissipating energy.

Earlier studies have reported that the electrostatic interactions between collagen and the HA, H-bonds and salt bridges, are the main key players for transferring loads among the constituents. The continuous deformation of such interactions, with a macro stick-slip motion, dampens the kinetic energy.[16,49] Therefore, we study the evolution of H-bonds and, in agreement with **Figure 7A**, we notice slight differences of the behaviors among the topologies. Under both compressive/tensile impulses (**Figure 7B**), we have a steady amount of the H-bonds only when $v_i$ equals 50 m·s$^{-1}$ and 100 m·s$^{-1}$, confirming a higher relaxation time. In contrast, when the input velocity increases, after a first transient phase, H-bonds converges at lower numbers pointing out an enhanced energy dampening due to a reduced capability to transfer loads. Moreover, we notice a difference between the loading conditions that is related to the number of H-bonds: when considering compressive impulses we have a drop that gives a higher dampening effect. This comparison gives the final discriminating factor between the outcomes from the two different loading conditions, which are otherwise quite similar. As for the *y*-/*z*-directions, we notice that the displacements due to loads along *x*-axis cannot be separated from the natural vibrations of the atoms, and therefore it is challenging to observe either the energy propagation or its dissipation along other directions in the compounds.

Our work presents a number of further implications: a number of studies have explored the behavior of collagenous materials, but they have mainly investigated the behavior of single molecules or the material at the macro scale.[18,44,47,50–53] This study aims at completing the results achieved in previous works,[16,29] offering a different perspective on the behavior of non-mineralized collagenous macro-tissues such as the tympanic membrane. The eardrum is a three-layer membrane, encompassing collagenous fibers in the middle layer with specific topologic arrangements. The tympanic membrane collects and filters the mechanical waves collected by the auricle, and transfers a selection of the vibration content to the ossicular chain.[54] While the macroscale behavior of the tympanic membrane collagen fibers has been studied, correlating the role of the different collagenous layers and fibers to specific functions (*e.g.*, radial fibers – main structural stability), a description of the mechanics of the micro/nanostructure of the eardrum is still missing.[55] Since we study the fibrillary level of the collagen, our description of the material could be embedded into a model of the material ultrastructure in order to investigate how the different conformation of the fibers, across the eardrum thickness, is related to the privileged transmission of the vibrations that occurs transversally to the median surface of the eardrum, while the mechanical waves are dissipated along the other directions. Moreover, since each area of the eardrum vibrates with different characteristic frequencies and amplitudes, a more detailed model of the biomechanics of the material may serve to link this peculiar behavior to macroscale observations.[56]

Bone is the other material that contributes to the hearing sense. In the middle ear, where sound transmission occurs, bone is the main constituent of three ossicles that connect the eardrum to the cochlea. The ossicles, instead, absolve their function of transmitting energy through their stiffness and topology, while energy dissipation is mainly performed by the soft tissues connected to bone structures.[57]

Bone is also the constitutive material of the temporal bone that enables osseous sound conduction to stimulate the inner ear through its viscoelastic behavior.[58]

Our outcomes may be helpful to both sound transmission and conduction applications. A first possibility is the development of a bioinspired structure that, mimicking the viscoelastic behavior of bone up to fibrillary level, can be used as a constitutive material for bone replacements in case of traumas, injuries or surgeries.[57,59–61] A second application relies on the study and prevention of specific pathologies that induce structural damages of both pure and mineralized collagenous materials (*e.g.*, fibrosis, Ehlers-Danlos syndrome, *osteogenesis imperfecta*).[62–64] The natural dissipative behavior of the material acts, indeed, as a passive mechanism to prevent the propagation of cracks and ultimate failure.[27,65]

**Conclusions**

We investigate the mechanical behavior of the bone at the molecular scale to understand the mechanics of the material upon impulsive compressive/tensile loads through an *in silico* model of mineralized fibrils. We observe the propagation and dissipation of the kinetic energy along the overlap and gap regions, unveiling interesting material dependencies on the mineralization percentage and the input velocity. Our model gives a detailed description of the deformation field at the nanoscale for both collagen and the HA. The achieved outcomes can be included in existing analytical models (*e.g.*, [43,66,67]) to perform sensitivity analyses in order to observe the mechanics of the bone-like material from a different perspective, namely the dissipative behavior upon transient loads. From the physiology standpoint, this study improves the understanding of bone-/collagenous-materials under traumas or diseases.

Moreover, since collagen and bone are at the base of the musculoskeletal system, a further understanding of the native tissue may help the development of replacement devices with different functions, from sound transmission or conduction to structural purposes.[60,68]

Finally, although the main focus of this study is about the effect of the mineralization on the mechanical behavior of bone, our results are not limited to tissue engineering applications. The development of bioinspired/biomimetic materials may also benefit from our research: dedicated artificial intelligence approaches[69] may, indeed, exploit a tuned viscoelastic behavior in materials/structures that, upon transient loads (*e.g.*, impact, fatigue), may show superior toughness and strength than traditional solutions, preventing structural failures or body traumas.[70]

**Methods**

**Topologies**
We employ the atomistic model designed and validated in previous works[16,29] to study collagen at different percentages of mineralization: 0%HA, namely non-mineralized collagen that model pure collagenous structures (*e.g.*, tendons), 20%HA and 40%HS, to investigate how the mineral component affects the mechanical properties (**Figure 1B**). Although the mineralization in the actual bone is higher than 40%, we decide not to study, and eventually fabricate, stiffer constructs to avoid brittleness and toughness issues, according to literature.[18] This model, however, is able to take into account the complex interactions among the chemical components that are often neglected, considering the intrafibrillar mineralization.[16,71–75] Nevertheless, in order to study the wave propagation along the structure, we chopped the bonds at the edges of the simulation box, leaving the periodic conditions only along the *y*- and *z*-directions.

**Force field**
We use a modified version of the CHARMM force field for all the simulations as done in previous works.[46,76,77] We slightly modify the code to include the hydroxyproline (HYP) that is an important block of the collagen sequence. We use a parameter set for HYP that derives from previous atomistic calculations that match quantum mechanics models.[78]

**Equilibration**
We process the topologies using the modified CHARMM force field in the molecular dynamics code implemented in the LAMMPS software.[79] We define two main groups in each topology: a first group – BD – that includes the ending atoms at the extremities of each topology and a second group – MB – which is composed of the rest of the atoms in the model. For each topology, we equilibrate all the atoms through a first step of NVE, fixing the momentum and the temperature (with Langevin thermostat at 10 K) of the MB for 100 ps. Afterwards, we heat the MB up to 310 K (room temperature) in 50 ps, and then we keep the temperature constant for other 50 ps. We minimize the energy keeping the MB fixed along the *y*- and *z*-directions and the temperature at 310 K with the Berendsen thermostat for 400 ps. Finally, all the atoms are equilibrated with the NVT at 310 K for 400 ps. We use the particle-particle particle-mesh solver with $10^{-5}$ kcal/mol-Angstrom accuracy to compute the long-range Coulombic interactions, while for the short-range ones we use the Leonard-Jones potential with global switching cut-offs set to 1 nm and 1.2 nm. We use a time step of 1 fs.

**Loading phase**
We constrain the atoms that belong to one end of the BD, and we apply an impulsive load to the opposite end. The loading conditions are either a tensile or a compressive impulsive displacement with period (*T*) equal to 10 ps and velocity ($v_i$) ranging from 50 to 1000 m·s$^{-1}$. The total observation time is 144 ps (**Figure 2**). The employment of impulsive loads, as discussed in previous works,[22,56] is able to deliver a broadband signal that encompasses a large number of frequencies.

We slice the MB in 50 parts and normalize the averaged displacements of the C$_\alpha$ atoms in the MB with respect to the load amplitude, to investigate the mechanical properties of the material in both the overlap and gap regions. The evolution of the displacement field gives the wave speeds in the two regions: $v_{OV}$ and $v_{GAP}$ for the overlap and gap respectively. Due to the expected differences between the two regions, we calculate the averaged wave speed ($v_{av}$) in the material as

$$v_{av} = \frac{v_{OV}T_{OV}+v_{GAP}T_{GAP}}{T_{OV}+T_{GAP}} = \frac{L_{OV}+L_{GAP}}{\frac{L_{OV}}{v_{OV}}+\frac{L_{GAP}}{v_{GAP}}} \qquad \text{Eq. 1}$$

in which $T_{OV}$, $T_{GAP}$ are the temporal intervals required by the mechanical wave to go through the overlap ($L_{OV}$) and gap ($L_{GAP}$) regions.

Taking advantage of the theory of vibrations, we correlate the wave speeds ($v$) with the Young's Moduli ($E$) and density ($\rho$) of the regions through the equation:[80]

$$v = \sqrt{\frac{E}{\rho}} \qquad \text{Eq. 2}$$

Moreover, we estimate the specific acoustic impedance $Z$ through its definition:

$$Z = \rho v \qquad \text{Eq. 3}$$

in order to evaluate the differences with the native bone tissue.[80]

We estimate the global stiffness of the material ($E_{av}$) by averaging the Young's Moduli of the two regions ($E_{OV}$, $E_{GAP}$) with the mass of such regions ($m_{OV}$, $m_{GAP}$):

$$E_{av} = \frac{E_{OV} m_{OV} + E_{GAP} m_{GAP}}{m_{OV} + m_{GAP}} \qquad \text{Eq. 4}$$

Finally, we model the dissipative behavior of the material by estimating the relaxation time ($\tau$) that describes the evolution of the kinetic energy – through $v_{Max}$, the maximum velocity of the wave front, fitted through the equation:

$$v_{Max}^2 \sim \exp(-t \cdot \tau^{-1}) \qquad \text{Eq. 5}$$

A further observation of the dampening phenomena is also carried out through the evolution of the H-bonds in the topologies through a dedicated code developed in house that elaborates the displacements of all the atoms with the original files of the structures.

**Theoretical models**

We compare our results with the outcomes from and two analytical models that correlate the Young's modulus ($E$) with the %HA.

The first model was developed by Halpin-Tsai[43] that describes the longitudinal modulus of a unidirectional plane, parallel to the platelet-reinforced composite as

$$E = E_m \frac{1 + A \times B \times \%HA}{1 - B \times \%HA} \qquad \text{Eq. 6}$$

in which $E_m$ is the modulus of the matrix, estimated as 0.5 GPa upon quasi-static loads in [16]. $A$ and $B$ are two constants given by

$$A = 2\gamma \quad , \quad B = \frac{\frac{E_p}{E_m} - 1}{\frac{E_p}{E_m} + A} \qquad \text{Eq. 7}$$

in which $\gamma \sim 30$ is the aspect ratio of the mineral, and $E_p = 100$ GPa is the Young's modulus of the mineral component.[43]

The second model was proposed by Gao, in which the elasticity of a nanocomposite is

$$E = \left[ \frac{4(1 - \%HA)}{\%HA^2 \times \gamma^2 \times G_p} + \frac{1}{\%HA \times E_p} \right]^{-1} \qquad \text{Eq. 8}$$

in which $G_p = 0.03$ GPa is the shear modulus of collagen.[67]

**Conflict of interest**
All the authors declare no conflict of interest.

**Acknowledgments**
This work was supported by the European Union's Horizon 2020 research and innovation program under the Marie Skłodowska-Curie grant agreement COLLHEAR No 794614. G.S.J. and M.J.B. acknowledges additional support from ONR (N000141612333) and AFOSR (FATE MURI FA9550-15-1-0514), as well as NIH U01HH4977, U01EB014976, and U01EB016422.

**Author contribution**
M.M., S.D. and M.J.B. designed the research. M.M. and G.S.J. implemented the model and analysis tools, carried out the simulations and collected the data. M.M, G.S.J., S.D. and M.J.B. analysed the results and wrote the paper.

**Supporting Information Available:** We provide a video that shows a qualitative depiction of the traveling and reflected waves. This material is available free of charge *via* the Internet at http://pubs.acs.org.

**Captions**

**Figure 1.** A. Hierarchical structure of the bone from the macroscale (skeletal tissue) to the atomistic scale (Tropocollagen and HA). B. Collagenous topologies with different mineralization percentages (0%, 20%, 40% of HA). In the lenses, we depict the tropocollagen and the HA crystals. These are oriented to align their *c*-axis with the fibril axis.

**Figure 2.** A: Mechanical model used in the study with the *D*-period fixed at one end, and loaded on the other end with either a compressive/tensile impulsive displacements. B: Impulsive load employed in the model with $T$ equal to 10 ps, $v_i$ ranging from 50 m·s$^{-1}$ to 1000 m·s$^{-1}$, and $t_f$ equal to 144 ps.

**Figure 3.** Mesh plots depicting the normalized displacements of the C$_\alpha$ atoms along the *x*-axis of the model as a function of the input velocity and mineralization percentage with both the push (Panel A) and pull (Panel B) loading conditions. This pictures help understanding the wave propagation (through the slopes of the displacements) and the energy dissipation through the displacement's fading. Moreover, the influence of the mineral can be noticed through the variation of the slope between the overlap and gap regions (we use dashed lines to represent the interface between the parts of the model).

**Figure 4.** Mesh plot depicting the normalized displacements of the C$_\alpha$ atoms of the 20%HA topology along the *x*-axis of the model as a function of the time. Traveling (black curved arrow) and reflected waves (red arrows) along the 20%HA compound with $v_i$ equals 50 m·s$^{-1}$. Reflected waves are generated every time the main traveling wave crosses the interface between the overlap and gap regions. From the energy balance, we estimate an average percentage loss of 7.55% with highest values for high %HA every time the traveling wave crosses the interface between the overlap and gap regions (black dashed line). An animation of this Figure is also given in the Supplementary Information. Note: the waves are not sinusoidal, the representation is purely qualitative to show a trend.

**Figure 5.** Wave speeds (A) and Young's Moduli (B) estimated using molecular dynamics simulations, by considering separately the outcomes for the Overlap and Gap regions, while C shows the averaged results as a function of the input velocity and the mineralization percentage. We observe how the wave speed and the Young's moduli decrease over the input velocity and how the mineralization significantly affects the propagation of the waves as well as stiffens the compounds.

**Figure 6**. Analytical models and results from atomistic simulations to correlate the Young's modulus and the %HA. Blue lines show the HT predictions, using as two different values of Young's moduli for the collagen matrices: $E_m = 0.5$ GPa for quasi static load (dashed line), and $E_m = 2.72$ GPa from the present work with impulsive loads (solid line). Star markers identify the outcomes from the impulsive load model across input frequency. The red solid line is a quadratic fit of the points shown with the star markers. The cyan solid line is presenting Gao's model, with which an earlier study[16] (blue circled markers) using quasi-static loads quite agree. The results from the current work show an underestimation of the Young's modulus with respect to the associated HT model (solid blue line), as also occurred in a previous work.[16] However, the outcomes are quite higher than the estimation made in [16] and Gao's model because the system is fast loaded, triggering the loading-rate behavior of the collagenous component.

**Figure 7.** Dissipation in mineralized collagen fibrils. A: relaxation time ($\tau$), along the *x*-axis, over input velocity ($v_i$). Average trend (red line) and standard deviation (shaded areas) by considering all the topologies with both the loading conditions. Independently from the loading conditions and the mineralization stage, the relaxation time has a monotonic decreasing trend, meaning great adaptability in dampening high-speed loads. B shows the evolution of the H-bonds over time under compressive and tensile loading conditions. We report the case of 20%HA, as representative, due to the similarities of the trends observed for the other topologies. After a first transient, the number of H-bonds remains steady only at low input velocities. In contrast, at 500 m·s$^{-1}$ and 1000 m·s$^{-1}$ we notice a convergence at lower numbers, giving an insight of the capability of the material to dampen energy. The observation of the H-bonds gives also the main discriminant of the behavior upon opposite loads with a higher dampening effect under pulling conditions.